\documentclass[aps,pre]{revtex4}
\usepackage{graphicx}
\begin{document}
\title{On the nature of long-range contributions to pair interactions between
charged colloids in two dimensions}
\author{Vladimir Lobaskin}
\affiliation{Max-Planck-Institut
f\"ur Polymerforschung, D-55128
Mainz, Germany}
\author{Matthias Brunner, Clemens Bechinger, Hans Hennig von Gr\"unberg} 
\affiliation{Fachbereich Physik, Universit\"at Konstanz, 78457 Konstanz, Germany }
\date{\today} 
\begin{abstract}
  We perform a detailed analysis of solutions of the inverse problem applied to
  experimentally measured two-dimensional radial distribution functions for
  highly charged latex dispersions. The experiments are carried out at high
  colloidal densities and under low-salt conditions. At the highest studied 
  densities, the extracted effective pair potentials contain long-range
  attractive part. At the same time, we find that for the best distribution
  functions available the range of stability of the solutions is limited by the
  nearest neighbour distance between the colloidal particles. Moreover, the
  measured pair distribution functions can be explained by purely repulsive
  pair potentials contained in the stable part of the solution. 
\end{abstract}
\maketitle

The machinery of statistical mechanics is designed to obtain information on
the structure of liquids from given interparticle interaction potentials.
When solving the inverse problem of statistical mechanics, one hopes to
find a unique interaction potential reproducing a measured distribution
function \cite{henderson74,rev:raja}. There are, however, always obstacles
such as limited range and finite accuracy of the measured distributions,
but also a number of numerical difficulties. No numerical procedure is
able to fit the reference distribution exactly. Therefore, in practice, the
inverse problem is always ill-posed and the uniqueness of the solution is
not guaranteed. The problem becomes even more complicated at high particle
densities where the spatial distributions are governed by packing effects.
In this case, a wide range of effective potentials is projected onto a very
tight space of radial distribution functions (rdf), thus making it
impossible to distinguish between the potentials of different shapes by
comparing the pair distributions only.

Another problem that occurs is particularly virulent in dense
suspensions of charged colloids. It is related to the fact that a
colloidal system is not a simple liquid with state-independent
pair-interactions, but rather a complex system in which the
colloidal interaction results from integrating out the
micro-ionic degrees of freedom
\cite{levin02,rev:vlachy,rev:loewen,reviewBelloni00}. As a result, a
description in terms of pair potentials becomes inadequate at higher
volume fractions when many-body interactions between the colloids come
into play. If this is the case, an inversion of the rdfs results in
density-dependent pair-potentials which contain contributions of the
many-body interactions and which are thus different from the true
(density-independent) pair-potentials acting between the particles.

In an attempt to explore the density dependence of the effective
interactions between charged colloids, we have recently made progress
in the experimental as well as the theoretical approach
\cite{brunner,klein}: (i) a much improved range and precision in
measuring rdfs at high colloidal number densities was achieved, (ii)
control of the colloidal densities became possible, and (iii) advanced
inversion tools were used. In our previous papers we have focused our
analysis on the behavior of the pair-potentials at short distances
where ambiguities due to the inversion procedure can be safely ruled
out. We have found that the pair potentials show a density-dependent
deviation from the expected Yukawa form which can best be explained in
terms of many-body interactions. We only mentioned that at larger
distances and high densities we have found attractive parts in the
pair-potential. The present paper now concentrates on the analysis of
this attraction.

\begin{figure}
\begin{center}
\vskip 0.2in
\includegraphics[width=0.75\textwidth]{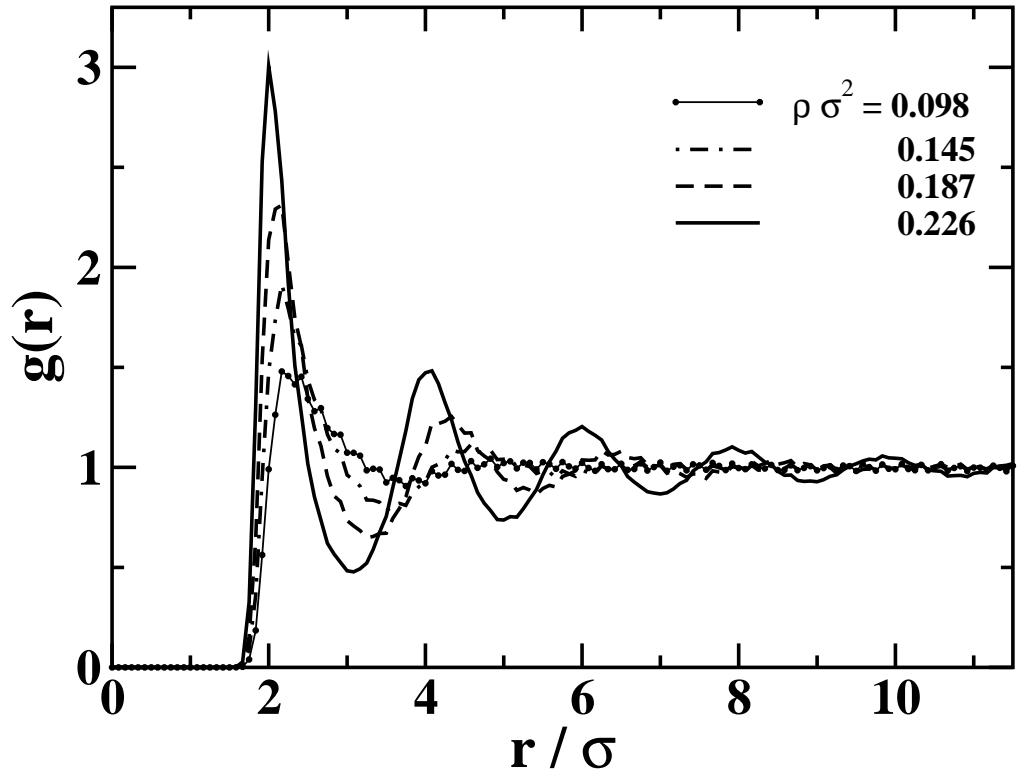}
\end{center}
\caption{
  Radial distribution functions for charged colloids in a 2D
  suspension at different colloidal densities, measured in a
  video-microscopy experiment.  $\sigma$ is the diameter of a colloidal
  sphere.}
\label{fig:1}
\end{figure}
From the beginning, we should make a distinction between the phenomenon 
we discuss here and the attraction observed between charged colloids
confined by two narrowly spaced glass plates ($<20\:\mu m$) or
located at the air-water interface
\cite{att1,carbajal3745,behrens05,cruz4203,Quesada,han03,ram03}. These
latter potentials were obtained at low densities for which different
inversion procedures gave almost identical results. Especially the
latest studies \cite{han03,ram03} show very clearly that pair-potentials
between colloids in weakly confined suspensions are purely repulsive,
while more strongly confined suspensions display attractions at the same
density; so confinement is obviously an essential requirement for the
observed attraction in those studies \cite{han03,ram03}. We did not
detect any similar confinement effects since the spacing between the two
glass plates of our sample cell was an order of magnitude larger than in
all the other experiments \cite{han03,ram03} ($>$ 200 $\mu m$).

In the present work, we consider a 2D charge stabilized colloidal
dispersion of $\sigma=2.4 \mu m$ diameter polystyrene sulfonate
particles taken at 2D packing fractions $\rho \sigma^2 $ ranging between
0.1 and 0.23, where $\rho$ is the particle number density in two
dimensions. More precisely, we should call our system a {\em quasi-2D}
system since the electric double layers around the colloids preserve
their three-dimensional character while the centers of the colloids are
effectively confined to a plane. The colloidal 2D rdf's were measured as
described in \cite{brunner,klein}, but with a higher spatial resolution
than in the previous works. As previously, an Ornstein-Zernike equation
(OZ)-based inversion routine with Percus-Yevick (PY) and
hypernetted-chain (HNC) closures as well as inverse Monte Carlo method
(IMC) were used to extract the effective pair potentials between the
colloids from the measured rdfs
\cite{LL95,LL97,lobaskin01,brunner,klein}.
\begin{figure}
\begin{center}
\vskip 0.2in
\includegraphics[width=\textwidth]{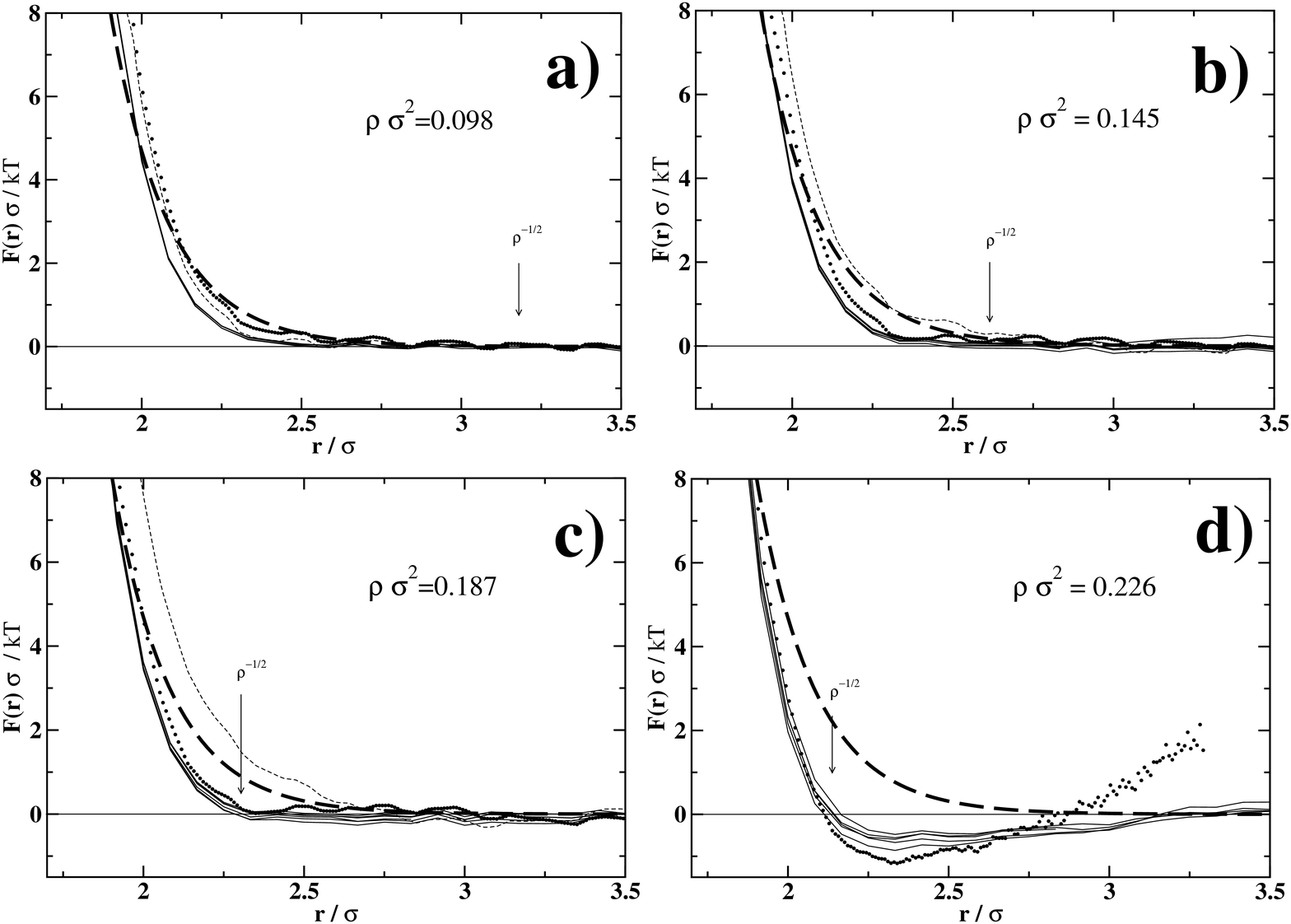}
\end{center}
\caption{
  Effective pair forces between two macroions in a 2D colloidal
  suspension at the indicated colloid densities, as obtained from
  experimentally measured rdfs using the inverse Monte Carlo method
  (thin solid lines) as well as integral equation scheme with HNC
  (dashed line) and PY (dotted line) closures and consequent numerical
  differentiation of the effective pair potentials.  To estimate the
  error, the IMC inversion has been carried out using different
  cut-off distances at $3 \sigma$, $5 \sigma$, $8 \sigma$ and $12.5
  \sigma$; each of these calculations is represented by one of the
  thin solid lines. A reference Yukawa potential is given as a thick
  dashed line.}
\label{fig:2}
\end{figure}
\begin{figure}
\begin{center}
%\vskip 0.2in
\includegraphics[width=0.5\textwidth]{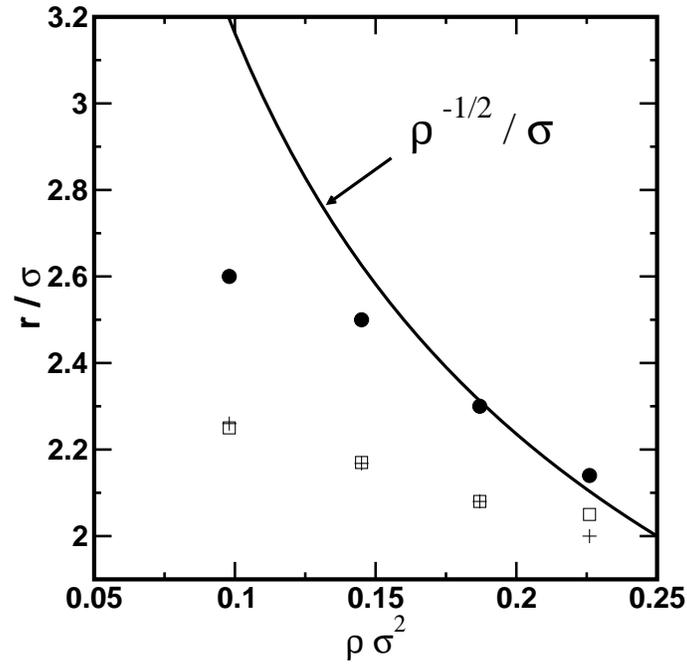}
\end{center}
\caption{
  The point of zero force (filled circles), the branching point (open
  squares) and the position of the first peak in $g(r)$ (crosses) vs
  the reduced densities, compared to the mean distance $d_{m}/\sigma =
  \frac{1} {\sqrt{\rho} \sigma}$ (solid line) between two colloids in
  a suspension at density $\rho$.
}
\label{fig:3}
\end{figure}
Figure~\ref{fig:1} shows the rdfs for the 2D latex dispersion. One can
easily observe that the position of the initial peak shifts slightly
towards smaller $r$ when the density is increased and that its height
grows with the concentration. The results of the inversion are
presented in Fig.~\ref{fig:2}. Since the potentials are defined by the
inversion procedure up to an additive constant and thus subject to
vertical shifts, it is more convenient to compare force curves, which
in Fig.~\ref{fig:2} are plotted for various concentrations, obtained
using different inversion methods. In the IMC method, the cut-off
radius within the inversion scheme (defined in
\cite{LL95,lobaskin01,klein}) has been varied in order to study its
effect on the final result. The collection of IMC curves in
Fig.~\ref{fig:2} thus gives an idea of the error produced by the
cut-off. A finite cut-off also implies that different lengths of the
rdf are used for the analysis. We thus model the experimental
situation where the range of accessible rdfs is shortened. For
comparison, we also plotted the best Yukawa fit for the IMC result at
the lowest density 0.098, denoted as the ''reference Yukawa
interaction'' in the following (pair-potential in units of $kT$:
$u(r)/kT = 35000\sigma\exp(-5r/\sigma)/r$).

The effective pair forces shown in Fig.~\ref{fig:2} display a very
steep repulsive part at short distances, whereas their long-range part
beyond $2.5 \sigma$ is close to zero. The potentials extracted from
the OZ-based method with PY and HNC closures differ form each other.
The HNC result approaches the reference Yukawa derivatives, while the
PY forces stay closer to the IMC data. At densities above 0.187 for
HNC no meaningful solution of the OZ based method could be found.
Some of the curves are slightly attractive at larger separations. The
depth of the minimum reaches $0.25 kT$ for the potential and $0.2
\sigma kT$ for the force. At the highest density, the uncertainty of
the result is fairly large in the long-range part. It is important to
note that the distances, at which the force reaches zero, correspond
roughly to the position of the main peak of the rdf at $r \leq
\rho^{-1/2}$ (arrows in Fig.~\ref{fig:2} indicate the mean
interparticle distance $\rho^{-1/2}$).  This may be seen in
Fig.~\ref{fig:3} where we compare the mean distance with (i) the
distance of the first peak in the rdf and (ii) the distance where the
force reaches zero.

The important new message of Fig. \ref{fig:2} is that the effective
pair-potential decomposes into two parts: a short-ranged and
cut-off-independent repulsive part and a long-ranged part, which can
show an attraction. This attraction depends sensitively on the chosen
cut-off. We call the cross-over point dividing the potential into a
cut-off-dependent and a cut-off-independent part the ''branching
point'' because beyond this point a noticeable branching of the
different IMC curves sets in.  Fig.~\ref{fig:3} demonstrates that the
position of this branching point scales with the density, and that it
is always at shorter distances than the point of zero force. The force
value, at which it starts, characterizes the sensitivity threshold of
the inversion procedure in use. We have to conclude that the
limitations of the inversion procedure allows us to make firm
statements only about the short-ranged repulsive parts of the
pair-potential, but not about the ''attractive'' forces observed at
larger distances. We stress that at distances smaller than the
branching point, the effective forces are clearly seen to be less
repulsive than expected for a Yukawa-like interaction, an effect which
increases with the concentration. This feature, observed also in 3D
suspensions \cite{jureEPL,jurejpcm,jureJCP}, is discussed in
\cite{brunner,klein,jureJCP} in terms of a macroion shielding effect
\cite{russ}; this result is obviously not affected by the uncertainty
of the inversion procedure.
\begin{figure}
\begin{center}
\vskip 0.6in
\includegraphics[width=0.7\textwidth]{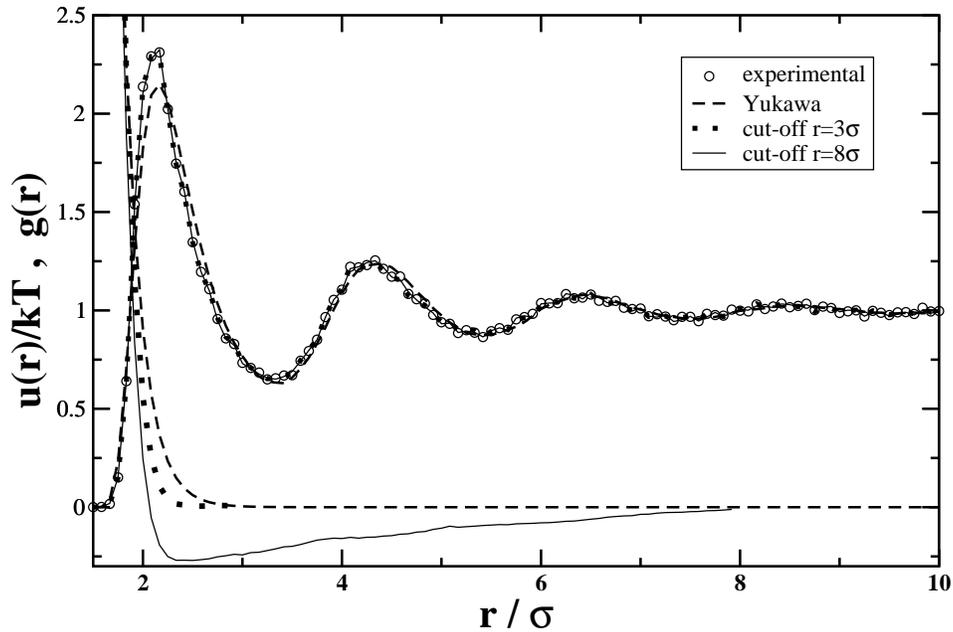}
\end{center}
\caption{
  Radial distribution functions and corresponding effective pair
  potentials for charged colloids in a 2D suspension at a colloid
  density of $\rho \sigma^2 = 0.187$. The potentials are obtained from
  experimental rdfs (circles) by means of the inverse Monte Carlo method
  using different cut-off distances, $r = 3 \sigma$ (dotted curve) and
  $r = 8 \sigma$ (solid curve). These two potentials lead to almost
  identical rdfs (solid and dotted rdf curves) that agree perfectly
  with the experimental rdf; their difference is comparable to the line
  thickness.  Also shown is the reference Yukawa potential and its
  corresponding rdf (thick dashed lines).}
\label{fig:4}
\end{figure}

The fact of branching manifests the apparent degeneration of the
solution to the direct problem, i.e., the calculation of the rdf. The
consequences of this degeneration of the problem at hand are further
illustrated by Fig.~\ref{fig:4}, which demonstrates the insensitivity
of the structure to the long-range part of the effective pair
potential. We show two potentials obtained from inverting the $g(r)$
measured at $\rho \sigma^2 = 0.187$, using a cut-off at $3 \sigma$ and
$8\sigma$, together with the rdfs generated from these potentials. To
facilitate comparison between the range of both the potentials and the
distribution functions, we present both quantities in the same plot.
While the potentials show a significant discrepancy, the difference
between calculated rdfs is smaller than the statistical uncertainty
(the sum of relative deviations form the reference rdf did not exceed
0.5\% in both cases) and show an excellent agreement with the
experimental data. The potential with the short cut-off turns to zero
at $r > 3\sigma$ while the long cut-off leads to a minimum of about
$-0.25kT$ at $r=2.5\sigma$. This corroborates our conclusions (pointed
out above) that the features of the effective pair interaction in a
distance regime beyond the first layer of neighboring particles cannot
be resolved for the present system. Certainly, these conclusions do
not apply to dilute systems where the effective potential is close to
the potential of mean force; then the inversion gives unambiguous
results \cite{behrens05,han03,ram03}.

One should note that for the observations made in this work, the
geometry of the system is essential. In all the mentioned quasi-2D
colloidal dispersions, only the colloidal particles are confined by
the external forces while the ionic clouds remain essentially
three-dimensional. A certain fraction of counterions escapes into the
bulk and thus does not participate in the in-layer screening, which
is confirmed by the extremely large Debye screening length (about 500
nm) obtained for these systems. As the double layers thicken, we have
a chance to probe their nonlinear parts and the many-body
contributions to the interparticle interaction \cite{russ}. For
comparison, in an unconstrained suspension of the same particles at
volume fraction of 0.1 (giving  a similar mean interparticle
distance), the deviation of the full electrostatic potential from its
far-field Debye-H\"uckel asymptote would be less than 5\% already at
$r=1.5\sigma$, while in our 2D system we still see significant
deviation from the Yukawa shape even at $r\approx 2 \sigma$. 

\begin{figure}
\begin{center}
\vskip 0.2in
\includegraphics[width=0.7\textwidth]{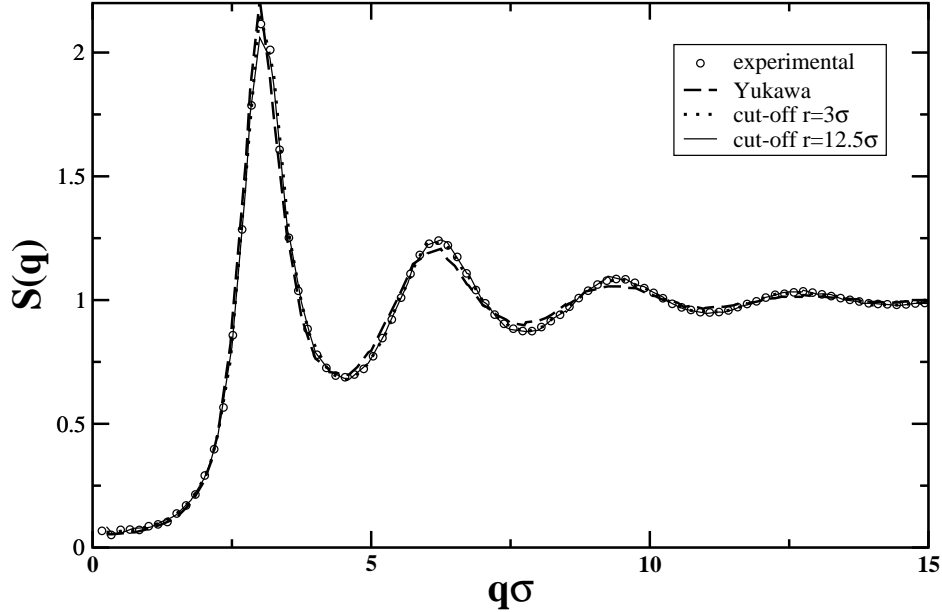}
\end{center}
\caption{
  Structure factors for charged colloids in a 2D suspension at a
  colloid density of $\rho \sigma^2 = 0.187$ as obtained using
  different effective potentials: IMC result with a cut-offs at $r = 3
  \sigma$ (dotted curve) and $r = 12.5 \sigma$ (solid curve), the
  reference Yukawa potential (thick dashed curve). The cut potentials
  improve the Yukawa result both at high q and in the limit $q
  \to 0$.}
\label{fig:5}
\end{figure}

Despite the apparent failure of our efforts to find the unique
solution of the inverse problem, this study still arrives at a useful
result. As it is seen in Fig. \ref{fig:4}, we were able to reproduce
the rdf to a very high accuracy using only the short-range part of the
effective potential. We compare in Fig. \ref{fig:4} the rdf obtained
using the reference Yukawa potential with that using the IMC result.
One can see that the model with the Yukawa potential predicts the rdf
notably worse. The deviation from the reference rdf in this case is
caused by the too strong repulsion at the mean interparticle distance.
Based on the comparison to the IMC curves that give excellent fits,
one can impose a simple correction on the reference Yukawa potential
to improve the agreement with the experimental rdf. The Yukawa
potential can be truncated at the position of the main rdf peak and
then shifted down so that the new potential is zero at and beyond this
distance. Since the position of the peak scales as it is shown in Fig.
\ref{fig:3}, the  new effective potential reflects the density change.
This idea of using a truncated Yukawa potential as a model for
colloidal pair-potentials has been tested in more detail in
\cite{jureJCP,jurejpcm}. 

It is also instructive to have a look at the osmotic coefficients of
our colloidal suspension. These coefficients have been obtained for
$\rho\sigma^{2} = 0.187$ from the virial equation, where we used the
various potentials discussed above. The IMC potential with a long
cut-off ($r = 12.5 \sigma$) gives 1.35 while the potential with the
shorter cut-off ($ r = 3 \sigma$) leads to a much larger osmotic
coefficient of 3.93, which is somewhat lower than that for the
reference Yukawa potential which is 4.48. The agreement between the
results of the latter two, on the one hand, and the large difference
to the result of the long cut-off potential, on the other hand, marks
a trend that does not match our observations made in Fig.~\ref{fig:4}
where the two cut potentials agree while showing differences to the
rdf of the reference Yukawa potential. As for the structure factor
(see Fig.~(\ref{fig:5})), one again observes the opposite trend: the
two cut potentials lead to almost identical results, which are clearly
different from the structure factor one obtains using the pure Yukawa
potential. This applies, especially, to the $q \to 0$ behavior. For
$S(q)|_{q \to 0} $ we find 0.074 and 0.066 for the potentials with the
longer and the shorter cut-offs, respectively, and 0.054 for the
Yukawa potential. Thus, although with the cut-off procedure 
we obtained improvement over the Yukawa potential in all studied
properties, there arises a potential dilemma: corrections of the
pair-potentials leading to improvements concerning the structure
of the liquid may spoil a possible agreement on the thermodynamic
side. In other words, the contribution of the many-body interactions
to the effective pair potential strongly depends on the property one
uses to detect these contributions \cite{louis02}. In view of this
fact, one could consider alternative ways to correct the effective
potential, for example, one which would correct not for the structure
(compressibility) but for the equation of state. 

We close with the statement that the best available structure data yet
do not provide sufficient accuracy for finding the unique solution of
the inverse problem for two-dimensional colloidal dispersions at high
densities.  Although we are able to deduce the pair potential at short
distances ($r<\rho^{-1/2}$), the limited accuracy of the rdfs do not
provide unambiguous information about the long-range part of the
pair-potential ($r>\rho^{-1/2}$). To narrow the range of the possible
solutions, we suggest applying an additional constraint on the
potential, i.e. to choose the potential that is equal to zero beyond
the first nearest neighbor layer. This cut-off procedure is justified
by the form of the pair potential, which is measured at low densities
where the many-body effects are minimal and thus is closest to the
true pair potential. We have shown that this choice produces
satisfactory pair distributions and improves description of
thermodynamic properties.

Authors thank David Grier for the valuable comments on the manuscript.
VL thanks the Swiss National Science Foundation for financial support
during his stay in the University of Fribourg.

\end{document}